\documentclass[sigconf,authorversion,nonacm,natbib=false]{acmart}

\AtBeginDocument{}

\RequirePackage[%
  datamodel=acmdatamodel,
  style=acmnumeric,
  defernumbers=true
]{biblatex}
\RequirePackage[nameinlink]{cleveref}
\RequirePackage{acronym}
\RequirePackage{svg}
\RequirePackage{xcolor}
\RequirePackage{listings}
\RequirePackage[htt]{hyphenat}

\DeclareSourcemap{
  \maps[datatype=bibtex]{
    \map[overwrite]{
      \step[fieldsource=skipbib, match=\regexp{true}, fieldset=options, fieldvalue={skipbib=true}]
    }
  }
}

\DeclareFieldFormat{url}{\url{\thefield{url}}}
\newbibmacro*{cfr:url}{\printfield{url}}
\DeclareCiteCommand{\citeurl}{%
    \boolfalse{citetracker}%
    \boolfalse{pagetracker}%
    \usebibmacro{prenote}%
}{\usebibmacro{cfr:url}}{\multicitedelim}{\usebibmacro{postnote}}

\newcommand\footnotecite[1]{\footnote{\citeurl{#1}}}

\definecolor{keywordcolor}{HTML}{0033B3}
\definecolor{stringcolor}{HTML}{067D17}
\definecolor{decoratorcolor}{HTML}{9E880D}
\definecolor{funccolor}{HTML}{00627A}
\definecolor{instancecolor}{HTML}{871094}

\lstdefinestyle{python}{
  language=Python,
  otherkeywords={async},
}

\lstdefinestyle{java}{
  language=Java,
  otherkeywords={var},
}

\begin{document}

\title{A Reference Architecture for Embedding Quantum Software into Enterprise Systems}

\author{Marc Uphues}
\email{marc.uphues@fh-muenster.de}
\orcid{0009-0005-3643-1210}
\affiliation{%
  \institution{Münster University of Applied Sciences}
  \city{Münster}
  \country{Germany}
}

\author{Sebastian Thöne}
\email{sebastian.thoene@fh-muenster.de}
\orcid{0009-0002-0812-8140}
\affiliation{%
  \institution{Münster University of Applied Sciences}
  \city{Münster}
  \country{Germany}
}

\author{Herbert Kuchen}
\email{kuchen@wi.uni-muenster.de}
\orcid{0000-0002-6057-3551}
\affiliation{%
  \institution{University of Münster}
  \city{Münster}
  \country{Germany}
}

\renewcommand{\shortauthors}{Marc Uphues, Sebastian Thöne, and Herbert Kuchen}

\begin{abstract}
Quantum computing promises a remarkable performance boost for certain applications, including computational intensive problems addressed by enterprise systems.
However, software architectures of enterprise systems must consider specific characteristics and quality attributes when collaborating with quantum computing services.
Hence, this paper presents a modular reference architecture for embedding quantum software into enterprise systems.
Its building blocks consist of loosely coupled and distributed services that implement both quantum-independent and quantum-specific tasks.
Although these services either depend on the business domain or the selected quantum algorithm, their orchestration forms a stable and reusable pipeline, specified as an executable BPMN model.
For demonstration and evaluation purposes, the proposed reference architecture is utilized in two case studies addressing combinatorial challenges from the field of operations research.
\end{abstract}

\maketitle

\acrodef{NISQ}{Noisy Intermediate-Scale Quantum}
\acrodef{VQA}{Variational Quantum Algorithm}
\acrodef{QAOA}{Quantum Approximate Optimization Algorithm}
\acrodef{QSE}{Quantum Software Engineering}
\acrodef{QCaaS}{Quantum-Computing-as-a-Service}
\acrodef{NSP}{Nurse Scheduling Problem}
\acrodef{QPL}{Quantum Programming Library}
\acrodefplural{QPL}[QPLs]{Quantum Programming Libraries}
\acrodef{API}{Application Programming Interface}
\acrodef{SOA}{Service-oriented Architecture}
\acrodef{UML}{Unified Modeling Language}
\acrodef{TSP}{Traveling Salesman Problem}
\acrodef{GPU}{Graphics Processing Unit}
\acrodef{TPU}{Tensor Processing Unit}
\acrodef{QPU}{Quantum Processing Unit}
\acrodef{BPMN}{Business Process Model and Notation}
\acrodef{DMN}{Decision Model and Notation}
\acrodef{XML}{Extensible Markup Language}
\acrodef{JSON}{JavaScript Object Notation}
\acrodef{TOSCA}{Topology and Orchestration Specification for Cloud Applications}
\acrodef{OpenQASM}{Open Quantum Assembly Language}

\section{Introduction}\label{sec:intro}

Enterprise systems are integrated software applications designed to support the business processes of an organization \cite{Bass.2022,Fernando.2023}.
They store relevant business information and execute data operations on business objects.
In some cases, enterprise systems also involve complex and computational intensive problems, e.g. for decision support, employee scheduling, or resource planning \cite{Fernando.2023}.
As quantum computing promises a remarkable performance boost for certain applications \cite{Acharya.2025}, it is tempting to leverage the advantages of quantum algorithms for addressing these combinatorial challenges.
In this field, one notable candidate is the \ac{QAOA} \cite{Farhi.2014}.
This quantum algorithm can be adapted to address several combinatorial challenges often seen in business-related contexts \cite{Brandhofer.2023,Kurowski.2023}.

Over the last century, quantum computing has transitioned from a purely experimental technology to a viable venture option.
For example, major cloud platforms now offer \ac{QCaaS} on a pay-per-use basis \cite{Leymann.2020}, such as IBM Quantum\footnotecite{IBMCorporation.2023b}, Amazon Braket\footnotecite{AmazonWebServicesInc..03.05.2024}, and Azure Quantum\footnotecite{MicrosoftCorporation.05.05.2024}.
Those platforms allow executing quantum software on quantum devices from various vendors via remote invocation \cite{Moguel.2022}.
This encourages organizations to consider quantum computing for their business needs, as it eliminates the necessity to invest in own hardware and know-how.

\begin{figure}[b]
    \centering
    \includesvg[width=0.45\textwidth,inkscapelatex=false]{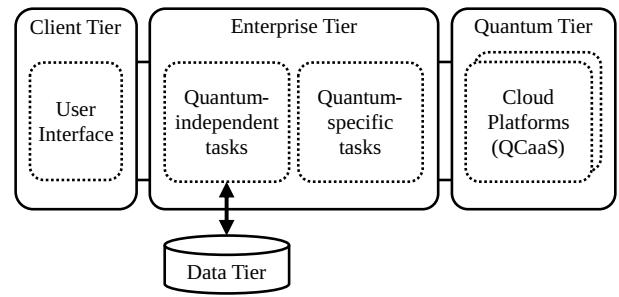}
    \caption{Architectural tiers of enterprise systems embedding quantum software}
    \Description{...}
    \label{fig:11_schematic-overview}
\end{figure}

To embed quantum software into enterprise systems, the enterprise and quantum tiers must be integrated accordingly \cite{PerezCastillo.2022} (see \autoref{fig:11_schematic-overview}).
This leads to hybrid quantum applications, which combine classical components for data processing and a quantum device through \ac{QCaaS} for computational tasks \cite{Weder.2021}.
The field of \ac{QSE} suggests the exploration of techniques for designing these hybrid classical-quantum systems \cite{Piattini.2020}, as they introduce several new challenges in the software engineering process \cite{Sodhi.2021,Stirbu.2023,Zappin.2025}.
When designing the necessary architecture, the software architect must identify and arrange additional components for the classical-quantum integration.
During the design process, the following questions are expected to arise:
\begin{itemize}
    \item Which quantum-specific tasks are inherent in the utilization of quantum algorithms through \ac{QCaaS}?
    \item How can a software architecture be designed that handles these tasks and integrates various specialized and potentially incompatible platforms?
    \item Which parts of that architectural design can be reused across various business problems, and which other components must be interchangeable depending on the business domain or selected quantum algorithm?
    \item Which further quality attributes should be incorporated by the software architectural design?
    \item What are appropriate platforms for realizing the resulting software architecture?
\end{itemize}

In this paper, we will provide answers to these questions and propose a modular reference architecture, which consists of a set of loosely coupled and distributed services.
Those can be adapted to any problem domain and quantum algorithm, while their orchestration remains as a stable and reusable pipeline.
This pipeline can be specified as an executable process model using the \ac{BPMN}.

The paper is organized as follows.
In \Cref{sec:problem}, we describe tasks inherent in quantum algorithms and establish requirements for an appropriate software architecture.
Further on in \Cref{sec:related-work}, we outline associated work dedicated to the software engineering of hybrid quantum applications.
The reference architecture is presented in \Cref{sec:solution}, while \Cref{sec:case-study} reports on two case studies we investigated for demonstration and evaluation purposes.
Eventually, we discuss our findings and state our conclusions in \Cref{sec:conclusion}.

\section{Problem Statement}\label{sec:problem}

The enterprise tier combines quantum-independent and quantum-specific tasks (see \autoref{fig:11_schematic-overview}).
Quantum-independent tasks are required regardless of the intended algorithmic solution (either classical or quantum).
These are common functionalities of enterprise systems, including routing, orchestration, as well as data aggregation and operations \cite{Fernando.2023}.
Quantum-specific tasks include data processing and mappings, as well as the generation of an algorithm-specific quantum circuit and its execution on a quantum device \cite{Weder.2022}.
Such tasks reoccur among current quantum algorithms, which allows for establishing an adaptable task sequence.
Since quantum-independent tasks have already been thoroughly researched \cite{Bass.2022,Fernando.2023}, we will focus on quantum-specific tasks inherent in the utilization of quantum algorithms.

Several challenges are encountered during the embedding of quantum software into conventional software systems \cite{Sodhi.2021,Stirbu.2023,Zappin.2025}.
Hence, best practices and standardized methods are needed for a proper integration.
Moreover, some characteristics are essential and certain quality attributes must be incorporated for embedding quantum software \cite{Sodhi.2021}.
We will examine the necessity of those, as it is crucial to design a suitable software architecture for an integrated enterprise system \cite{Fernando.2023}.

\subsection{Reoccurring Quantum-Specific Tasks}\label{sec:problem:tasks}

The following quantum-specific tasks reoccur among current quantum algorithms \cite{Weder.2022}:
\begin{enumerate}
    \item Circuit generation per problem instance,
    \item circuit execution on the desired quantum device,
    \item circuit refinement of the quantum circuit,
    \item measurements interpretation of resulting quantum states and probability distributions.
\end{enumerate}
These are determined by the quantum algorithm at hand.
\acp{QPL}, such as Qiskit\footnotecite{IBMCorporation.2023b} and PennyLane\footnotecite{Xanadu.2023b}, provide reusable software components for circuit-related tasks.
In the following, we will give a brief explanation for each of these.

\subsubsection{Circuit Generation}

Quantum algorithms are based on quantum circuits, which require translation of input data from classical to quantum domains \cite{Sodhi.2021,Stirbu.2023}.
Concrete problem instances must be translated into qubits and quantum gates, resulting in a distinct circuit per problem instance \cite{Rojo.2021}.
For this purpose, \acp{QPL} offer functions to map certain mathematical models, such as graphs and matrices, to the corresponding quantum circuits \cite{Weder.2022}.
However, domain-specific data must still be mapped by the enterprise tier to align with these models.

\subsubsection{Circuit Refinement}

Depending on the quantum algorithm, the circuit may require additional refinement.
For instance,\linebreak parametrized quantum algorithms need further optimization of the generated circuit based on the given problem instance \cite{Weder.2022}.
In the case of the \ac{QAOA}, which relies on an iterative optimization cycle, this includes approximating optimal parameters for the gates of the generated circuit \cite{Farhi.2014}.

\subsubsection{Circuit Execution}

As of today, a quantum circuit is typically executed on a specified quantum device through \ac{QCaaS} \cite{Leymann.2020}.
For this, a quantum device needs to be selected from a range of predefined vendors.
\acp{QPL} provide parametrized interfaces to initiate and monitor the circuit execution on the selected cloud platform \cite{Weder.2022}.
After a specified number of executions (\textit{shots}), the resulting state measurements and probability distributions are returned \cite{Weder.2022}.

\subsubsection{Measurements Interpretation}

The obtained probability distributions of quantum states are interpreted to derive a solution for the problem at hand \cite{Weder.2022}.
This process requires a mapping of these states back to the originating business domain and its concepts.
Notably, \acp{QPL} lack the functionality for such reverse mapping.

\subsection{Quality Requirements for Software Architectures}\label{sec:problem:requirements}

When embedding quantum software into enterprise systems, several characteristics must be considered during the design phase of a suitable software architecture.
It should incorporate the following quality attributes, based on current industry standards \cite{InternationalOrganizationforStandardization.2023}:
\begin{enumerate}
    \item Modularity,
    \item Maintainability,
    \item Multi-Platform Compatibility,
    \item Availability,
    \item Scalability,
    \item Adaptability,
    \item Reusability.
\end{enumerate}
In the following, we provide a brief explanation of the stated quality attributes and their necessity for embedding quantum software into enterprise systems.

\subsubsection{Modularity}

The software architecture should consist of components following the principle of single responsibility \cite{Martin.2014}.
This principle mandates that each component should only be responsible for a single aspect of the system's functionality.
As a result, quantum-independent and quantum-specific components designed with this principle in mind can be reused across business domains without unnecessary dependencies.

\subsubsection{Maintainability}

Although \acp{QPL} aim to simplify the process of constructing quantum circuits by providing supportive tools and code abstractions, they are still highly specialized for quantum-specific tasks that developers seem to struggle with \cite{Zappin.2025}.
Hence, the resulting components of the software architecture must be loosely coupled to enhance flexibility and facilitate independent development.
This allows developers with different specializations to concentrate on their specific areas of expertise.

\subsubsection{Multi-Platform Compatibility}

Components dedicated to\linebreak quantum-independent and quantum-specific tasks are usually implemented using different platforms.
Quantum-independent components are built in mainstream programming languages, such as Java and C\#.
They rely on the robustness of established and stable platforms, such as the \textit{Spring}\footnotecite{VMwareInc..2023} framework.
In contrast, quantum-specific components are built within scientific and computational environments required by \acp{QPL}.
Thus, the software architecture must ensure compatibility for heterogeneous platforms to support these diverse environments.

\subsubsection{Availability}

Quantum circuits are executed on quantum devices via remote invocation (\ac{QCaaS}).
These devices are managed by third-party vendors and may not be consistently available \cite{Sodhi.2021}.
This can lead to potential delays, preventing the immediate return of circuit execution results \cite{Stirbu.2023}.
In addition, the execution duration of some quantum-specific tasks can be uncertain \cite{Stirbu.2023}.
As a result, the architecture should account for the limited availability of quantum devices, manage potential delays, and prevent blocking behavior due to the uncertainty in task execution duration.

\subsubsection{Scalability}

The enterprise tier includes both quantum-\linebreak independent and quantum-specific tasks, which can vary in complexity and computational load.
If these differing computational demands are not properly managed, it can lead to inefficient resource utilization or the creation of bottlenecks.
Therefore, it is recommended to design components of the enterprise tier with scalability in mind, ensuring that the system can handle varying computational needs.

\subsubsection{Adaptability}

Quantum computing is expected to experience ongoing advancements.
Emerging quantum algorithms are anticipated, and cloud platforms are expected to provide access to even more advanced quantum devices.
To leverage the latest advancements and further mitigate the risk of potential vendor lock-in, the resulting architecture should be designed with corresponding adaptability in mind, ensuring it can easily adapt to new quantum-specific components, platforms, and devices \cite{Stirbu.2023}.

\subsubsection{Reusability}

Serving as a design reference, the software architecture should be reusable across different business domains.
This verifies its general applicability and allows for practical application of quantum computing to a broader range of use cases, which further encourages organizations to explore quantum computing for their respective business needs.

\section{Related Work}\label{sec:related-work}

Over the last years, researchers have explored the aspects of hybrid quantum applications in the field of \ac{QSE}.
In the following, we outline work related to the design and orchestration of software architectures for these kinds of systems.

\textcite{Wild.2020} propose an extension for the \ac{TOSCA}.
This extension provides the ability to automate the deployment and orchestration of quantum-specific services.
For this, the quantum algorithm is encased in its own service, allowing for reusability across various software systems.
However, the quantum circuit generation and its execution are tied strongly together.
This may cause reduced adaptability of the resulting service and the impression of potential vendor lock-in.
To enable adaptation to both emerging quantum algorithms and advanced quantum devices independently, it seems beneficial to separate these quantum-specific tasks.

Within the scope of orchestration, \textcite{Weder.2021} propose a workflow-based approach to orchestrate classical and quantum-specific tasks.
With the help of a provisioning engine, all software artifacts of a quantum application are deployed accordingly.
Moreover, a workflow engine initiates the sequence of activities determined by the quantum application.
The conclusion of their work states the verification of this approach by an initial prototype based on the workflow platform \textit{Camunda}\footnotecite{CamundaServicesGmbH.29.01.2025b}.
Yet, their work lacks a demonstration showcasing the prototype's application in practical scenarios.
Such a demonstration would help organizations to assess the feasibility and potential applications of quantum computing.

\textcite{Rojo.2021} present an implementation of different quantum microservices for solving the \ac{NSP}.
Their results state the already mentioned inconvenience of tightly coupled quantum code, devices, and cloud platforms.
However, the work lacks a structured approach to task and service orchestration.
Additionally, the proposed software architecture consists merely of the quantum microservice, lacking a broader architectural perspective for contextual integration into enterprise systems.
To demonstrate feasibility for organizations, it is crucial to showcase the proper integration of enterprise systems and quantum software.

In the context of quantum microservices, \textcite{Moguel.2023} introduce a continuous deployment architecture to automate the software delivery process of quantum microservices.
This approach intends to generalize the process of designing and implementing a quantum microservice by automation.
Its goal is to reduce the complexity introduced by quantum-specific components.
For this, the process of code generation and deployment is automated by utilizing a DevOps-pipeline.
Yet, their work lacks a practical demonstration showcasing this approach within enterprise systems.

\section{Solution}\label{sec:solution}

We propose a reference architecture for enterprise systems embedding quantum software.
Following a top-down approach, the overall design consists of two layers: The \textit{quantum orchestration pipeline} and a modular service architecture responsible for handling the pipeline's tasks.

First, the \textit{quantum orchestration pipeline} decomposes the sequence flow inherent in the utilization of quantum algorithms and organizes reoccurring tasks accordingly.
It includes the previously identified quantum-specific tasks, as well as additional ones introduced to support fulfillment of the suggested quality attributes.
The objective of this model is to provide a reusable pipeline.

Second, a collection of loosely coupled and distributed services is intended to handle the specified tasks from the proposed pipeline.
An orchestrator determines the task sequence and controls the flow by assessing the pipeline model.
For the latter, the orchestrator distributes commands to the individual services via asynchronous communication.
Additionally, it tracks the state of each pipeline instance.
The objective of this architecture is to ensure the necessary characteristics and to incorporate the suggested quality attributes.

Additionally, we propose a pattern for algorithm selection and invocation inspired by the strategy design pattern \cite{Gamma.1995}.
This pattern allows for consideration of various algorithms at runtime (either classical or quantum).
The adoption of this pattern is a novel approach towards the dynamic selection of a suitable quantum algorithm with adaptability and extensibility in mind.

\begin{figure*}
    \centering
    \includesvg[width=0.99\textwidth,inkscapelatex=false]{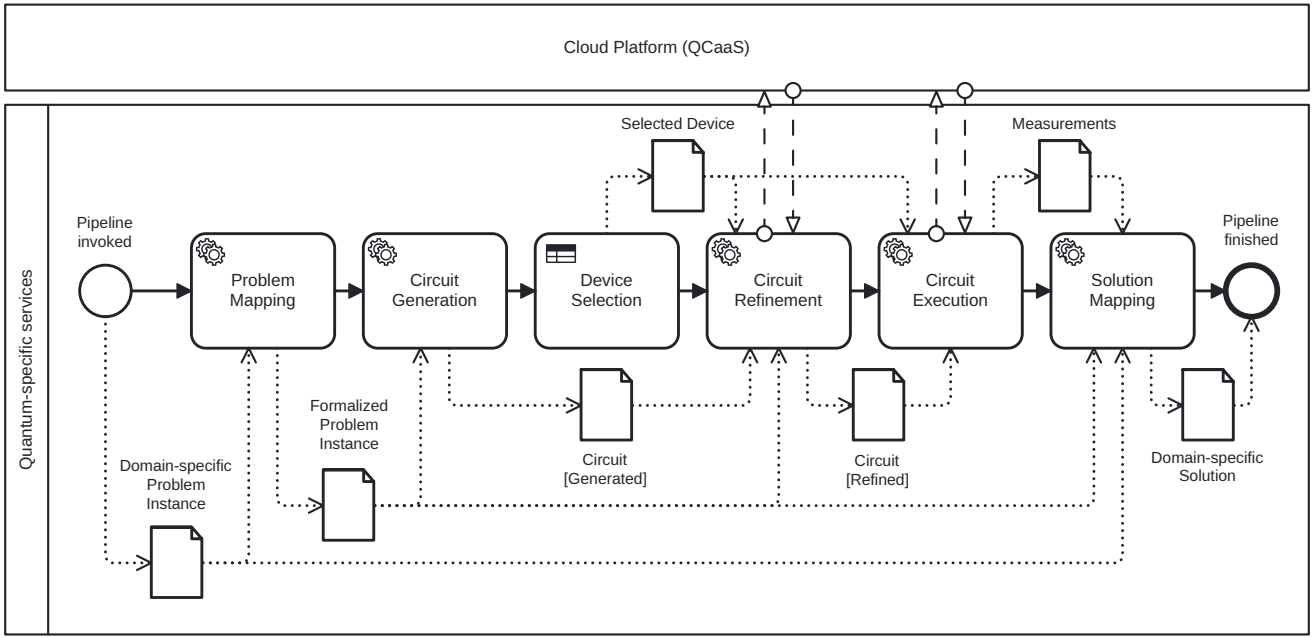}
    \caption{The quantum orchestration pipeline in \ac{BPMN}}
    \Description{...}
    \label{fig:21_pipeline}
\end{figure*}

\subsection{Quantum Orchestration Pipeline}\label{sec:solution:pipeline}

The \textit{quantum orchestration pipeline} seen in \autoref{fig:21_pipeline} arranges the reoccurring quantum-specific tasks previously identified.
For enhancing modularity, we separate the problem mapping of domain-specific data to formal mathematical models from the circuit generation.
In this regard, we also introduce the task of quantum device selection.
Moreover, the task of measurements interpretation is replaced by solution mapping, emphasizing its alignment with the problem mapping task.
Ultimately, the \textit{quantum orchestration pipeline} specifies the following quantum-specific tasks:
\begin{enumerate}
    \item Problem mapping of domain-specific data,
    \item circuit generation per problem instance,
    \item device selection for circuit execution,
    \item circuit refinement of the generated quantum circuit,
    \item circuit execution on a quantum device,
    \item solution mapping aligned with problem mapping.
\end{enumerate}
Although the implementation of these tasks often depends on the selected quantum algorithm at hand, the proposed orchestration pipeline can be reused, as long as the implementing services are kept interchangeable.
This flexibility allows for the adaptation to different algorithms without requiring significant modifications to the overall structure.
We explain the flow of the pipeline as follows.

\subsubsection{Problem Mapping}

When the process gets invoked by an incoming request, a domain-specific problem instance is provided.
These domain-specific data are mapped to align with specific mathematical models, as \acp{QPL} provide functions to map these to corresponding quantum circuits.
However, the tasks for mapping and circuit generation are separated to further maintain modularity.
Based on this, the service task produces a formalized problem instance, which is passed on to the circuit generation task.

\subsubsection{Circuit Generation}

Within this service task, a quantum circuit is generated based on the formalized problem instance.
In general, the given input data must be translated into qubits and quantum gates \cite{Rojo.2021}.
The explicit implementation for this process is determined by the used quantum algorithm.
The outcome of this process is a generated circuit, which is forwarded to the circuit refinement task.
For this data transfer, a cross-platform representation for quantum circuits is required.
As such, the \ac{OpenQASM} \cite{Cross.2021} can be utilized.

\subsubsection{Device Selection}

To execute the quantum circuit, a suitable quantum device needs to be chosen from a range of predefined vendors.
It can be taken using a set of business rules defined with the help of a decision table.
In this context, multiple factors can be considered, such as the device's number of qubits, its availability, as well as service charges and execution costs.
It is important to note that the chosen quantum device must fulfill certain requirements inherent in the generated circuit, such as the number of qubits.
Therefore, this step is intended to be executed after the circuit is initially generated.
The selected device is considered by the circuit refinement and circuit execution tasks.

\subsubsection{Circuit Refinement}

As stated previously, a generated quantum circuit may require additional refinement.
The need for this service task heavily depends on the used quantum algorithm and it determines the implementation of this task.
Parametrized quantum algorithms, such as the \ac{QAOA}, need further optimization of the circuit based on the given problem instance.
Subsequently, the refined circuit is transferred to the circuit execution task.

\subsubsection{Circuit Execution}

After the quantum circuit is generated and refined, it is executed on the selected quantum device through \ac{QCaaS}.
For this, \acp{QPL} provide suitable client functionality \cite{Weder.2022}.
After the execution has finished and a returning message has been received, the resulting quantum state measurements are passed on to the solution mapping task.

\subsubsection{Solution Mapping}

The obtained quantum state measurements are mapped in relation to the problem mapping.
To accomplish this, both the domain-specific problem instance and the formalized problem instance are required in this process.
Eventually, the solution mapping task preserves the domain-specific solution, which is returned to the invoking process for further processing.

\subsection{Deployment View and Building Blocks}\label{sec:solution:architecture}

\begin{figure}
    \centering
    \includesvg[width=0.485\textwidth,inkscapelatex=false]{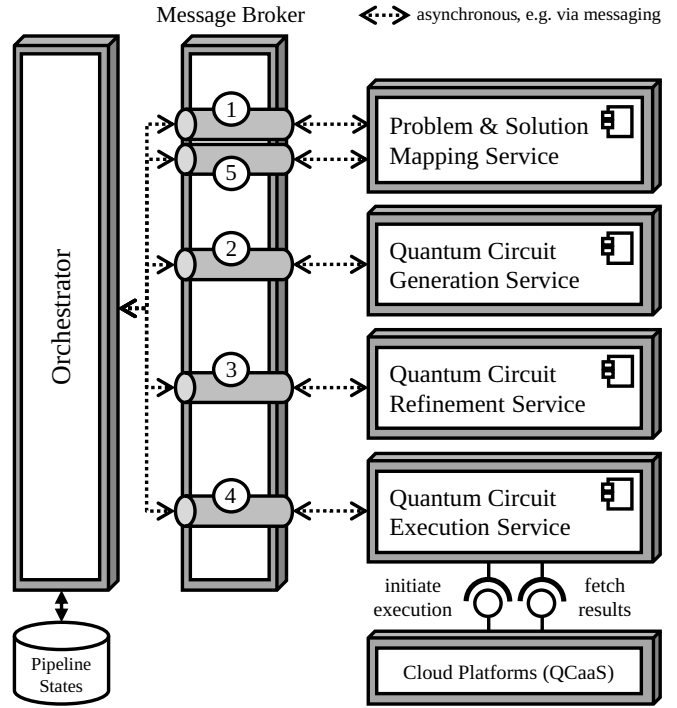}
    \caption{Deployment view of the proposed reference architecture}
    \Description{...}
    \label{fig:22_services}
\end{figure}

\autoref{fig:22_services} shows the deployment view of the proposed reference architecture.
Its building blocks consist of an orchestrator, a message broker, as well as several distributed services dedicated to handle quantum-specific tasks.

The orchestrator ensures that each task is initiated in accordance with the sequence flow, maintaining the logical order for correct process execution.
For controlling the flow, it tracks the state of each pipeline instance and examines the defined sequence to identify any subsequent tasks upon their completion.
To initiate the handling of tasks, it instructs the individual services via the message broker.

The message broker decouples the orchestrator from the task-handling services.
It achieves this through asynchronous communication, meaning that the individual components can send and receive messages at different times without waiting for immediate responses.
As certain quantum-specific tasks might take an unpredictable amount of time to execute, the prevention of blocking behavior caused by this is essential to ensure overall availability.

The deployment view seen in \autoref{fig:22_services} includes several services, each dedicated to handle one of the pipeline's tasks.
Some services' implementations are determined by the quantum algorithm at hand, including those of circuit generation and circuit refinement.
The problem and solution mapping service heavily depends on the business domain, as domain-specific data needs to be mapped accordingly.
Finally, the implementation of the circuit execution is independent of both, business domain and quantum algorithm, enabling its reuse across various scenarios, along with the orchestrator and message broker.
Furthermore, the services are loosely coupled, which facilitates multi-platform compatibility and scalability, ensuring each service can operate within its required and optimal environment.
This also promotes maintainability, as developers with different specializations can focus on their specific areas of expertise.
Additionally, this simplifies the adaptation to enhancements for various components, including mapping techniques, circuit generation implementations, and refinement methods.

We outline key interactions of the architecture in \autoref{fig:22_services} with the numbers labeled (1) to (5) as follows:

\begin{enumerate}
    \item At first, the mapping service is orchestrated to map the given domain-specific data to a mathematical model.
    \item Then, the generation service generates a quantum circuit based on the given problem instance and implemented quantum algorithm.
    \item If the resulting circuit needs further refinement, the refinement service is orchestrated to optimize it, depending on the quantum algorithm.
    \item After that, the execution service is instructed to initiate the execution of the quantum circuit and fetches the resulting quantum state measurements via \ac{QCaaS}.
    \item At the end, the mapping service applies appropriate functions to map the obtained results back into domain-specific data formats.
\end{enumerate}

\newpage
\subsection{Extension by Multiple Strategy Support}\label{sec:solution:strategy}

\begin{figure}
    \centering
    \includesvg[width=0.49\textwidth,inkscapelatex=false]{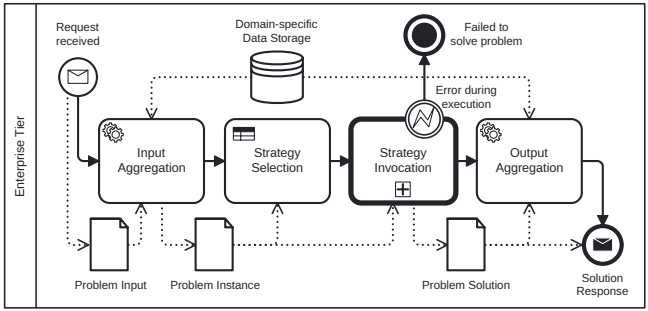}
    \caption{The \textit{strategy decision pattern} in \ac{BPMN}}
    \Description{...}
    \label{fig:23_strategy-decision}
\end{figure}

With the \textit{strategy decision pattern}, an enterprise system can dynamically select an appropriate solution strategy (either classical or quantum) for a given problem instance, depending on relevant factors.
For example, a quantum algorithm may not be suitable for smaller problem sizes, whereas classical algorithms may be inefficient in certain circumstances.
The \textit{strategy decision pattern} seen in \autoref{fig:23_strategy-decision} consists of the following tasks for business problems:
\begin{enumerate}
    \item Input aggregation of domain-specific data,
    \item strategy selection per problem instance,
    \item strategy invocation for the given problem instance,
    \item output aggregation of domain-specific data.
\end{enumerate}
In the following sections, we explain the process and each task of the \textit{strategy decision pattern} in more detail.

\subsubsection{Input Aggregation}

Initially, an incoming request is received containing the problem input.
The input data usually consist of multiple domain-specific entities.
In addition, parameters contained in the input can control the outcome of the solution process.
For example, in the case of optimization problems, the input contains a range of constraints to satisfy.
The task is able to retrieve additional data from the domain-specific data storage for aggregation.

\subsubsection{Strategy Selection}

After the input aggregation, a decision is taken on the strategy to employ for addressing the problem instance.
This decision can be made using business rules defined in a decision table.
For instance, the problem's size could affect the strategy selection, as strategies based on quantum algorithms may be more suitable for larger problems than smaller ones.
The outcome of the strategy selection task includes a reference to a corresponding strategy process definition.
This is passed on to the strategy invocation task.

\subsubsection{Strategy Invocation}

The associated strategy is invoked as a sub-process, which is responsible for providing a solution.
For instance, the proposed \textit{quantum orchestration pipeline} can be utilized to solve the given problem using a quantum algorithm.
Nonetheless, the specific manner in which the process achieves a solution is beyond the scope of the \textit{strategy decision pattern}.
As each strategy may interpret the given problem differently, its mapping to algorithm-specific formats falls in the sub-process.
At the end, the strategy process must produce a domain-specific problem solution.

\subsubsection{Output Aggregation}

In the last step, the resulting problem solution is aggregated with additional domain-specific data.
The need for this service task depends on the business domain and expected outcome, therefore being optional.

\section{Case Studies}\label{sec:case-study}

For demonstration and evaluation purposes, we present two case studies exemplifying our proposals\footnote{\url{https://github.com/mu-qse/quantum-task-orchestration}}.
In each of these, we implement a typical enterprise system which embeds quantum software for solving a common combinatorial challenge from the field of operations research.
For their implementation, we utilize the proposed reference architecture (see \Cref{sec:solution}).

As a first case study, we select the well-known \ac{NSP} \cite{Miller.1976}.
It involves assigning employees to shifts in a way that satisfies multiple constraints.
To show the practical usability of our solution, we adapt this to a typical call center environment, where employees are also called \textit{agents}.
The goal is to create an optimal schedule that meets the organization's needs, while adhering to specific constraints.

As a second case study, we choose the knapsack problem \cite{Lawler.1977}.
It comprises the selection of items to maximize the total value of these selected items, while ensuring that their combined weight does not exceed a given capacity.
We adapt this to a common logistic scenario in which a cargo ship must be loaded with various containers, each differing in weight and value.
Given the ship's limited capacity, the goal is to determine an optimal composition that maximizes the total value without exceeding the capacity constraint.

Ultimately, our objective is to demonstrate the feasibility of embedding quantum software into enterprise systems.
Furthermore, we evaluate and discuss the systems with respect to the characteristics and quality attributes stated in \Cref{sec:problem}.

\subsection{Preparations and Setup}

To orchestrate the individual services, we select the orchestration platform \textit{Camunda}\footnotecite{CamundaServicesGmbH.29.01.2025b}.
It provides adaptable process automation, supporting established standards for modeling and orchestrating business processes, such as \ac{BPMN}, and is particularly well-suited for orchestrating the tasks of the \textit{quantum orchestration pipeline} and \textit{strategy decision pattern}, as we can utilize our proposed \ac{BPMN}-models (\autoref{fig:21_pipeline}, \autoref{fig:23_strategy-decision}) without much effort.
Moreover, the integrated workflow engine \textit{Zeebe}\footnotecite{CamundaServicesGmbH.29.01.2025c} promotes message-driven communication and loose coupling among distributed services, as required by our proposed reference architecture (\autoref{fig:22_services}).
For receiving events, services register themselves as workers via the \textit{Zeebe Client}.
This client is a supplementary software component provided through a library, implemented in multiple programming languages, including Java and Python, to ensure compatibility with various platforms for service workers.
Once a service is registered, it is able to fetch and process assigned tasks from the workflow engine.
For the assignment, each task is associated with a specific type, which defines the corresponding worker responsible for its receipt and execution.
This type-based assignment ensures that tasks are routed to the appropriate service workers.
Once the task completes, the worker sends its results back to the orchestrator.

For the implementation of domain-specific services, including those of problem and solution mapping (see \autoref{fig:21_pipeline}), as well as input and output aggregation (see \autoref{fig:23_strategy-decision}), we select the \textit{Spring} framework.
It is an established and robust framework for building enterprise applications in Java, therefore well-suited for our use case, as it promotes modularization and loose coupling through aspect-oriented programming and dependency injection.

To implement quantum-specific services inherent in the \textit{quantum orchestration pipeline}, we select a Python-based environment for utilizing the Qiskit framework.
It provides a comprehensive set of tools for generating, optimizing, and executing quantum circuits on simulators and quantum devices.
The framework is well-suited for our application, as it offers modules for generating the quantum circuits to solve the chosen combinatorial optimization challenges.

We deploy the orchestration platform and the workflow engine on a remote host running Docker\footnotecite{DockerInc..2023}.
The other services of the proposed reference architecture are deployed locally to showcase the capability for distributed processing.
To transfer data between the individual services, we employ the \ac{JSON}.
Moreover, we utilize the \ac{OpenQASM} \cite{Cross.2021} for transferring quantum circuits between tasks.

\subsection{Call Center Agent Scheduling}

At first, we import the models of the \textit{quantum orchestration pipeline} and \textit{strategy decision pattern} into the \textit{Camunda Modeler}.
As already mentioned, the workflow engine utilities type-based assignment of tasks and workers.
For each task, a custom job type must be set to associate it with the implementing services.
These types are defined for the tasks of \autoref{fig:21_pipeline} as follows:

\begin{enumerate}
    \item \texttt{\small scheduling\_qaoa\_problem-mapping},
    \item \texttt{\small scheduling\_qaoa\_circuit-generation},
    \item \texttt{\small quantum\_device-selection},
    \item \texttt{\small scheduling\_qaoa\_circuit-refinement},
    \item \texttt{\small quantum\_circuit-execution},
    \item \texttt{\small scheduling\_qaoa\_solution-mapping}.
\end{enumerate}

When implementing the individual services dedicated to handling the models' tasks, the services must be associated with the corresponding type.
The Zeebe Client library provides Java and Python-annotations for declaring functions as corresponding entry points for workers (see \autoref{code:java:problem-mapping}).

\begin{lstlisting}[
    style=java,
    caption=Entry point for the problem mapping in Java,
    label=code:java:problem-mapping,
]
<@{\color{decoratorcolor}@JobWorker}@>(
    type = "scheduling_qaoa_problem-mapping", 
    autoComplete = false)
public void problemMapping(
    final JobClient jobClient,
    final ActivatedJob job,
    <@{\color{decoratorcolor}@Variable@>(name = "problem") 
        final ScheduleProblemDto problem) {
    final var problemGraph = 
        this.<@{\color{instancecolor} problemMappingService@>.map(problem);
    jobClient.newCompleteCommand(job)
        .variable("problemGraph", problemGraph)
        .send();
}
\end{lstlisting}

Further on, we describe implementations of the resulting services for each of the tasks in more detail.

\subsubsection{Graph-based Problem Mapping}

\begin{figure}[h]
    \centering
    \includesvg[width=0.45\textwidth,inkscapelatex=false]{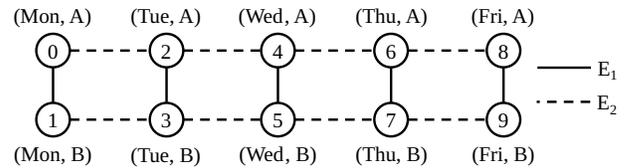}
    \caption{Representation of possible shift-agent assignments as vertices connected by the constraints $E_1$ and $E_2$}
    \Description{...}
    \label{fig:52_scheduling-graph}
\end{figure}

\acp{QPL} lack functionality for mapping domain-specific data to mathematical models, as mentioned in \Cref{sec:problem}.
Therefore, a custom implementation is needed to map the agent scheduling problem appropriately.
The problem can be described using an undirected graph.
For explanatory purposes, we will assume that two agents must be assigned to five shifts and certain constraints apply in this regard.
\autoref{fig:52_scheduling-graph} shows a visualization of possible combinations for this scenario.
The indexed vertices in this graph show the possible shift-agent assignments, while the edges between these vertices represent the following constraints $E_1$ and $E_2$:
\begin{itemize}
    \item $E_1$: Exactly one agent must be assigned per shift.
    \item $E_2$: No consecutive shifts may be assigned to the same agent. This constraint prevents an agent from being assigned to two consecutive shifts.
\end{itemize}
This graph-based mapping allows to frame the problem of agent scheduling as a max-cut problem, which can be solved by the adoption of \ac{QAOA} \cite{Farhi.2014}.

\subsubsection{QAOA-Circuit Generation}

For circuit generation, we utilize the module \texttt{\small QAOAAnsatz}\footnotecite{IBMCorporation.02.02.2025} from the Qiskit library.
Each vertex in the constructed graph (and therefore each possible shift-agent assignment) is being mapped to a qubit, while the edges (the constraints to satisfy) are encoded into the problem's objective function.
As a result, each bit within the bitstring of the quantum states corresponds to its respective vertex within the graph.

\subsubsection{Device Selection}

The selection of an appropriate device can be achieved using a specific set of business rules.
Several factors can be included in this set, such as the device's qubits and device availability.
For our case study, we realize a service with a list of predefined simulators and quantum devices to consider.
Since our focus is not on the definition of such business rules, the selected device is manually set to a locally executed simulator instance.

\subsubsection{QAOA-Circuit Refinement}

The process of circuit refinement is essential for the functioning of the \ac{QAOA}.
Within this task, the parameters associated with the quantum gates are iteratively optimized to improve the resulting quantum states produced by the \ac{QAOA}-specific circuit \cite{Farhi.2014}.
To achieve this, we utilize a local instance of the built-in simulator from the Qiskit library.
The resulting circuit is specifically optimized for the given problem instance.

\subsubsection{Circuit Execution}

For the execution of resulting circuits, a corresponding endpoint of the desired cloud platform is called.
The Qiskit library handles this tasks autonomously.
The input data required for the execution initiation include the quantum circuit itself, the information which device to run the circuit on, as well as the number of shots the circuit should be executed.
In our example, the measurements obtained from 1000 shots are collected, and the resulting probability distributions are summarized accordingly.

\subsubsection{Graph-based Solution Mapping}

For obtaining a solution, an appropriate function is implemented to interpret the results from the returned quantum state measurements.
The function takes the bitstring with the highest occurrence, indicated by the peak probability in the resulting distribution.
This bitstring corresponds to the solution that best satisfies the constraints and requirements of the problem instance.
As previously stated, each bit within the bitstring of the quantum states maps directly to a corresponding vertex from the indexed vertices of the graph.
Therefore, each bit represents a shift-agent combination that is either included or excluded from the solution.
A bit in state $1$ indicates that the associated combination is included in the derived solution, while state $0$ implies that the combination is excluded from the solution.
For instance, the bitstrings $1001100110$ and $0110011001$ represent valid solutions for the explanatory scenario of five shifts and two agents.
Eventually, the bits of the solution bitstring are mapped to specific entity instances by their index.

\subsection{Cargo Ship Container Composition}

For the second case study, we also import the proposed models into \textit{Camunda}, and a custom job type is defined for each of the pipeline tasks (\autoref{fig:21_pipeline}) as follows:

\begin{enumerate}
    \item \texttt{\small knapsack\_qaoa\_problem-mapping},
    \item \texttt{\small knapsack\_qaoa\_circuit-generation},
    \item \texttt{\small quantum\_device-selection},
    \item \texttt{\small knapsack\_qaoa\_circuit-refinement},
    \item \texttt{\small quantum\_circuit-execution},
    \item \texttt{\small knapsack\_qaoa\_solution-mapping}.
\end{enumerate}

As stated previously, the tasks of device selection and circuit execution are independent of the problem domain and quantum algorithm.
Therefore, we reuse their definitions and implementations from our first case study.

\subsubsection{Problem Mapping}

The implementation for problem mapping relies on a community-driven extension to Qiskit\footnotecite{QiskitOptimizationDevelopmentTeam.2024}.
It offers components dedicated to the solution of various optimization problems, including the knapsack problem.
As a result, the circuit generation task expects a list of values and weights of the selectable items, as well as the maximum capacity.

\subsubsection{QAOA-Circuit Generation}

Similar to our first case study, we use the module \texttt{\small QAOAAnsatz} provided by the Qiskit library.
In addition, we utilize the knapsack class from the Qiskit Optimization extension.
With the help of this class, a corresponding objective function can be derived by the given list of values and weights, constrained by the given capacity.
The resulting qubits include so-called \textit{slack qubits}, which are not actively engaged in quantum operations at certain times during the circuit execution (\textit{slack intervals}) \cite{Smith.2023}.

\subsubsection{Device Selection}

The selection of an appropriate device is not bound to a specific business domain or quantum algorithm, as previously mentioned.
Consequently, we reuse the implementation from our first case study for this task.

\subsubsection{QAOA-Circuit Refinement}

The process of circuit refinement is crucial for the \ac{QAOA}, as pointed out previously.
While the resulting \ac{QAOA}-circuit is specifically optimized for the given instance of the knapsack problem, the implementation for this task is similar to the first case study.

\subsubsection{Circuit Execution}

As the execution of a circuit is independent of the business domain, we can again reuse the respective implementation from our first case study.

\subsubsection{Solution Mapping}

As in our first study, an appropriate function is implemented to interpret the returned quantum state measurements in accordance with the problem mapping.
The bitstring with the highest probability represents the solution that optimally satisfies the constraints and requirements of the problem instance.
It contains $n$ bits (starting from the most significant bit), representing the given items to consider for the composition.
A bit in the state $1$ indicates that the item is included in the composition, while the state $0$ implies that the item is excluded.
Additionally, the resulting bitstrings contain several slack bits that are not associated to the solution \cite{Smith.2023}.
To determine the list of domain-specific entities representing the solution, the function considers the indices of the solution bitstring and the given list by the problem mapping task.

\subsection{Evaluation of the Results}

The case studies we conducted demonstrate the feasibility of embedding quantum software into enterprise systems by utilizing our proposed reference architecture.
All services of the architecture were implemented, while being orchestrated by the orchestration platform and the workflow engine, in accordance with \ac{BPMN}-models of the \textit{quantum orchestration pipeline} and \textit{strategy decision pattern}.

To evaluate the realized systems, we examine their characteristics and quality attributes (see \Cref{sec:problem}).
Each of the following aspects is rated on a scale from $1$ to $5$, based on the evaluation outcomes.
A score of $1$ indicates \emph{Not Fulfilled}, $3$ represents \emph{Partially Fulfilled and Partially Not Fulfilled in equal measure}, and a score of $5$ signifies \emph{Fully Fulfilled}.

\subsubsection{Modularity}

The resulting systems are handling the identified quantum-independent and quantum-specific tasks in accordance with the defined models.
Each of the tasks is implemented by a corresponding service worker.
Algorithm-independent services, such as those for device selection and circuit execution, can be reused across different problem scenarios as shown.
This also applies for the \textit{Camunda} platform, realizing the orchestrator and message broker respectively.
Algorithm-specific services, including those for circuit generation and circuit refinement, require an individual implementation per problem scenario.
However, they concentrate on their associated task within the business domain.
In addition, the individual services are testable in isolation, without requiring mocks of unrelated and unnecessary dependencies.
For instance, the services dedicated for input/output aggregation and problem/solution mapping were tested without requiring quantum-specific dependencies.
Hence, we infer high modularity of the software architecture (score of $5$).

\subsubsection{Maintainability}

During the realization of the case studies, we were able to implement the individual services independently and without requiring the other services to exist.
For instance, the services for problem mapping and circuit generation tasks were realized first, while the other services did not exist yet.
In addition, after all services were deployed, we were able to release new versions these without affecting the overall system.
However, the respective code for deriving the objective function of \ac{QAOA} is duplicated across the services for circuit generation and circuit refinement, as we were not able to transfer this function between the respective services.
Moreover, we also investigated a migration of the circuit execution from the existing Qiskit implementation to the PennyLane library.
Although both libraries support \ac{OpenQASM} for circuit representation, we encountered difficulties executing the circuit using the PennyLane implementation.
In the end, we derive a medium level of maintainability (score of $3$).

\subsubsection{Multi-Platform Compatibility}

The individual services of the resulting architectures are realized with different programming languages, frameworks, and environments.
For instance, the quantum-independent services were implemented with the \textit{Spring} framework written in Java, while the quantum-specific services were realized in a Python-based environment required by the Qiskit framework.
Additionally, we assume that the chosen environments are interchangeable, as long as they support the defined interfaces and data formats.
Moreover, the architecture is distributed across hosts with different operating systems for the orchestration platform, the workflow engine, and the individual services.
Therefore, we rate the support for heterogeneous platforms as high (score of $5$).

\subsubsection{Availability}

The implemented services register themselves as workers via the Zeebe Client.
As per type-based assignment, multiple services can register for a specific task.
If one service instance is unavailable, the Zeebe workflow engine chooses another instance for fulfillment and routes the task accordingly, without affecting the overall system's availability.
For instance, we deployed multiple instances of the circuit refinement service across different hosts.
This distribution was necessary due to the high computational load associated with circuit optimization, which caused these systems to become unresponsive. 
Moreover, the task assignment operates in an asynchronous and message-driven manner, which ensures that tasks are assigned without waiting for their immediate response.
Considering these aspects, we rate the availability of the resulted systems as high (score of $5$).

\subsubsection{Scalability}

The services are implemented in a stateless manner, which allows for horizontal scaling by adding more handling service instances for certain tasks.
In addition, the services are able to independently utilize vertical scaling through resource allocation per deployment.
This also applies to the orchestration platform and the workflow engine.
However, the scaling requires manual intervention to deploy additional service instances or adjust resource allocation.
Hence, we derive a reasonably high level of scalability (score of $4$).

\subsubsection{Adaptability}

By applying the \textit{strategy decision pattern} (\autoref{fig:23_strategy-decision}), the system is able to consider multiple solution strategies for a given problem.
These utilize different quantum algorithms, \acp{QPL}, and cloud platforms.
Moreover, the individual services are interchangeable within the strategy.
Therefore, we infer a high adaptability (score of $5$).

\subsubsection{Reusability}
In both case studies, we reused the proposed models for orchestration, requiring only minor adjustments to the type-based service assignment for the individual use cases.
Moreover, the distributed service architecture was applied to both resulting systems, each of which implemented the task-handling services independently.
However, we did not investigate the outcome when employing different algorithms for the chosen problems.
Consequently, we rate the reusability as rather high (score of $4$).

\subsection{Findings of the Case Studies}

The organization of tasks enables the resulting systems to achieve a high level of modularity by clearly stating the tasks' responsibilities and interfaces.
Each resulting service is dedicated to a single task of the \textit{quantum orchestration pipeline}, which ensures development and testing independence.
This also allows for reuse of certain services across business domains.
The loose coupling enables support for various platforms and environments required by quantum-independent and quantum-specific services alike.
Additionally, the utilization of asynchronous communication allows the components to operate without waiting for immediate response, ensuring high availability of the overall system.
Furthermore, the individual services are interchangeable, meaning that enhanced implementations can be deployed easily for adaption to new advancements. 
Besides that, the \textit{strategy decision pattern} enhances extensibility by enabling adaptation to new strategies.

We experienced several challenges regarding \acp{QPL}, confirming the findings of \textcite{Zappin.2025}.
For instance, the services of circuit generation and refinement are more related to each other than initially thought, resulting in certain code duplications across these services.
Additionally, generating and executing circuits across different libraries appears to be challenging.
Despite the existence of standards for circuit representation and transfer, the libraries seem to interpret the data in diverging ways.
For these reasons, it might be more practical to introduce a single service dedicated to quantum-specific tasks, as proposed by \textcite{Rojo.2021}.
However, this would lead to tighter coupling, which restricts the independent scalability of the corresponding services.
By separating the circuit generation and refinement, we were able to scale these accordingly.
Moreover, implementing a single service for quantum-specific tasks reduces reusability, as certain tasks are not bound to a specific quantum algorithm, such as the device selection and circuit execution.
With the introduction of individual services, we were able to reuse their implementations across different business domains.

\section{Conclusion}\label{sec:conclusion}

This paper addresses certain challenges introduced by the embedding of quantum software into enterprise systems.
To streamline the design process, we propose a reference architecture that maintains modularity and multi-platform compatibility.
By following the principle of single responsibility \cite{Martin.2014} even for quantum-specific tasks, this novel approach allows for independent development, individual scaling, adaptability and reusability of distributed quantum-specific services.
The \textit{Camunda} platform acts as a suitable realization platform, supporting the integration of these quality attributes.
Additionally, the \textit{strategy decision pattern} enables the consideration of multiple algorithmic solutions during runtime, allowing for extensibility and adaptability to advancements in quantum computing.

We applied the proposed \textit{quantum orchestration pipeline} in different business domains, each requiring its own set of task-handling services.
Regardless of their implementations, the orchestration of those services remained stable throughout these use cases.
However, additional case studies from various business domains could be explored to enhance the general applicability of the orchestration models.
This would also provide demonstrations to utilize quantum algorithms in different business domains.
Furthermore, the pipeline could be further refined by assessing a broader range of quantum algorithms.
Yet, \acp{QPL} only provide a selected range of quantum algorithms and lack functionality for converting domain-specific data into the required input formats for quantum algorithms.
This results in an inefficient and potentially error-prone implementation process.
New approaches could be explored to automatically derive these necessary mappings from existing entity definitions.

Maintaining modularity and multi-platform compatibility is essential when embedding quantum software into enterprise systems.
Our case studies demonstrate that a modular and loosely coupled service architecture is well-suited for the incorporation of these quality attributes.
Hence, we encourage software architects to follow a distributed approach using task-specific services to facilitate multi-platform compatibility, development independence, as well as scalability, adaptability, and reusability on a per-service level.


\printbibliography


\end{document}